\documentclass[twocolumn,longbibliography,prl,superscriptaddress,reprint]{revtex4-1}
\usepackage{amssymb}
\usepackage{graphicx}
\usepackage{dsfont}
\usepackage{amsmath}
\usepackage{bm}
\usepackage{times}
\usepackage{color}
\usepackage[applemac]{inputenc}
\usepackage[nice]{nicefrac}
\usepackage{float}
\usepackage[colorinlistoftodos]{todonotes}
\usepackage{ esint }
\usepackage{xcolor}

\PassOptionsToPackage{normalem}{ulem}
\usepackage{ulem}

\graphicspath{{converted_graphics/}}

\begin{document}

\title{von Neumann entropy and the entropy production of a damped harmonic oscillator}

\begin{abstract}
In this paper we analyze the entropy and entropy production of a non-isolated quantum system described within the quantum Brownian motion framework. This is a very general and paradigmatic framework for describing non-isolated quantum systems and can be used in any kind of coupling regime. We start by considering the application of von Neumann entropy to an arbitrarily damped  quantum system making use of its reduced density operator. We argue that this application is formally valid and develop a path integral method  to evaluate that quantity analytically. We apply this technique to a harmonic oscillator in contact with a heat bath and obtain an exact form for its entropy. Then we study the entropy production of this system and enlighten important characteristics of its thermodynamical behavior on the pure quantum realm and also address their transition to the classical limit.
\end{abstract}

\author{G. A. Weiderpass}
\altaffiliation[Present Address:]{ Department of Physics, The University of Chicago, 5720 South Ellis Avenue, Chicago, IL 60637}
\affiliation{Instituto de  F\'\i sica Gleb Wataghin, Universidade Estadual de Campinas, 13083-859, Campinas, SP, Brazil}
\author{A. O. Caldeira}
\affiliation{Instituto de  F\'\i sica Gleb Wataghin, Universidade Estadual de Campinas, 13083-859, Campinas, SP, Brazil}

\maketitle

One of the most celebrated contributions of Ludwig Boltzmann to statistical mechanics is the well-known  expression of the entropy of a system as the logarithm of the number of its accessible microstates. This revolutionary prescription enables us to obtain thermodynamical properties from purely statistical considerations, and has furnished us with what is now one of the most important tools of physics. Nevertheless, the application of Boltzmann entropy is not an easy task in most problems since obtaining the number of microstates accessible to a system under specific conditions may not be a trivial matter. It is generally more useful to use the {\it {von Neumann}} representation of the statistical entropy which can be written in terms of the density operator of the system, $\rho$, as
\begin{align} \label{von-Neumann}
S=-k_{B} \mathrm{tr} \rho \ln \rho.
\end{align}
This representation is particularly useful when we are dealing with a system subject to a certain number of constraints imposed by the external world.

Even though the applications of Boltzmann and von Neumann entropies have been quite successful over the years, there are situations in which the equilibrium state of a system (or a particular subsystem of the whole universe, to be more precise) does not result from the maximization  procedure as applied to the standard forms of the  entropy functions related to the subsystem observables only. This has been the source of many controversies and generated a vast literature on attempts to obtain generalized entropies which, in the appropriate limit, would converge to the von Neumann or Boltzmann entropies (see below). 

One of the most emblematic situations where the maximization of (\ref{von-Neumann}) might not lead to the correct equilibrium density operator comes from the theory of open quantum systems. In this theory, a system of interest is coupled bilinearly to a thermal reservoir which is modelled by a set of non-interacting harmonic oscillators. This is the well-known model for quantum Brownian motion which is the paradigm of open quantum systems in the scientific literature \cite{caldeira1,caldeira2,caldeira3,Weiss,Breuer}. The Hamiltonian of this model is given by
\begin{align} \label{HU}
\begin{aligned}
H = \frac{p^2}{2 m} + V(q)+\sum_{j}^{\infty} \Big[ \frac{p_{j}^2}{2 m_{j}} + \frac{m_{j} \omega_{j}^2}{2} \Big(q_{j} -\frac{C_{j}}{m_{j} \omega_{j}^2 }q \Big)^2\Big].
\end{aligned}
\end{align}
The theory also gives us a a specific prescription for the spectral function which is widely known in the literature and thus left for the supplementary material.

Therefore, if we consider the system and the reservoir together (which we call the universe), the maximization of the entropy subject to the appropriate constraints gives us the Gibbs state, $\rho=Z^{-1} \exp\{-\beta H\}$ with the Hamiltonian $H$ given by (\ref{HU}). Now, to obtain the equilibrium state of the system of interest, namely its reduced density operator, we must trace out the bath degrees of freedom from the universe density operator, which yields
\begin{align} \label{rho eq}
\rho_S(x,y,\beta)\equiv \textrm{tr}_R \rho =\frac{1}{Z}  \int d \textbf{R} \langle x, \textbf{R}|  e^{-\beta H} | y, \textbf{R} \rangle,
\end{align}
where $Z$ is the partition function evaluated with the Hamiltonian (\ref{HU}). 

It is well-known that this operator is not the Gibbs operator obtained with the Hamiltonian of the system of interest neglecting its coupling to the environment, except in the classical limit (high temperatures or $\hbar \rightarrow 0$) \cite{caldeira3,Weiss,Breuer,Durao,Oliveira}. The latter is obtained when  we use the system Hamiltonian $H_S=p^2/2m+ V(q)$ and  maximize the von Neumann entropy subject to the constraints $E_S=\textrm{tr} \rho_S H_S$ and $\textrm{tr} \rho_S=1$.

Generalized entropies were expected to be able to provide the correct density operators when maximized subject to the appropriate constraints. However, even though these generalized entropies are mathematically sound and good measures of the lack of information about the systems, they have failed to provided the correct density operator. The most representative cases are the Tsallis and R\'{e}nyi entropies \cite{Tsallis1,Renyi,Tirnakli,Moyano,Tsallis2}. It has been shown that the extremization of these two  entropies for open quantum systems does not lead to the correct density operator \cite{Durao}.

In this paper, we argue that the von Neumann entropy, expressed as a functional of the reduced density operator of a subsystem (our system of interest), is still the most adequate measure of its entropy, regardless of the fact that this equilibrium state may or may not be obtained directly from a maximization procedure applied to that very form of the entropy subject to general constraints. Before we proceed, let us touch upon some crucial issues regarding our main goal.

Firstly, why bother about associating an entropy for the system of interest once the partition  system-plus-environment does not necessarily obey the standard conditions  to safely define an entropy function? In other words, the interaction between system and reservoir cannot be neglected as usual. Although this is a pertinent question, one should bear in mind that despite this, there is a thermal dependence on ensemble averages of observables of the system of interest only. It is an inescapable conclusion therefore that a thermodynamical approach is indeed necessary to describe its equilibrium state. As an example we could mention the equilibrium variance of the position and momentum operators of an arbitrarily damped quantum harmonic oscillator which, by the fluctuation-dissipation theorem \cite{Foster,Zwanzig,LeBellac},  reads
\begin{align}
\begin{aligned} \label{p2 and q2}
    \langle q^2 \rangle &=\frac{2\hbar \gamma }{m\pi}\int\limits_{0}^{\infty}d\nu \coth{\frac{\hbar \nu}{2 k_BT}}\frac{\nu}{(\nu^2-\omega_0^2)^2+4\gamma^2 \nu^2} ,\\
\langle p^2 \rangle &=\frac{2\hbar \gamma m}{\pi}\int\limits_{0}^{\infty}d\nu \coth{\frac{\hbar \nu}{2 k_BT}}\frac{\nu^3}{(\nu^2-\omega_0^2)^2+4\gamma^2 \nu^2} .
\end{aligned}
\end{align}

Secondly, why choose the specific von Neumann form for the entropy? In order to answer this question let us consider a system which is very weakly coupled to the reservoir. Then, the time evolution of the whole universe may be described by \cite{Breuer}
\begin{align} \label{Born-Markov}
\rho (t) \approx \rho_S(t) \otimes \rho_R(t),
\end{align}
where $\rho_R(t)=\rho_R^{(eq)}$ is the equilibrium density operator of the reservoir ignoring its interaction with the system of interest. In other words, the interaction is not strong enough to disturb the equilibrium state of $R$. This is the well-known Born-Markov approximation which leads to many different master equations, the so-called rotating wave approximation being one of the most popular among them. Those equations are widely used in many different areas like optomechanics, photonics, spectroscopy, to name just a few \cite{Breuer,Mukamel,Fox}.

As we know that the von Neumann entropy of the whole universe leads to its correct density operator, we may apply it to (\ref{Born-Markov}). Thus, we obtain
\begin{align} \label{SU}
S_{U} = S_S + S_R,
\end{align}
with $S_S=-k_{B}\textrm{tr}_S \rho_S\ln \rho_S$ and $S_R=-k_{B}\textrm{tr}_R\rho_R \ln \rho_R$. This shows that in the weak coupling regime, which is the most ubiquitous case, the use of the von Neumann entropy of the system of interest as a functional of its reduced density operator is well established. The real difficulty appears when we try to apply (\ref{von-Neumann}) to situations where the coupling between system and environment does not allow us to employ the Born-Markov ansatz (\ref{Born-Markov}).  

Therefore, let us boldly assume that the decomposition  
(\ref{Born-Markov}) is still valid, but with $\rho_S(t)\, \textrm{and}\, \rho_R(t)$ being, respectively, the reduced density operators of the system and environment ($\rho_R \equiv \textrm{tr}_S \rho$). These two reduced density operators can be exactly evaluated for a quadratic system as the one described by (\ref{HU}). The non-trivial issue here is to prove that the decomposition (\ref{Born-Markov}) is, at least, approximately valid. We believe a possible case of success would be the example of quantum Brownian motion where a single particle is coupled to a much larger system. Although we recognize it has to be explicitly shown, we shall assume it as our fundamental hypothesis for the time being.

The third point has to do with the entropy production of this system. The entropy production is simply the total variation of the entropy of the universe. As the variation of entropy of the reservoir is given by the total heat it absorbs divided by the temperature, using (\ref{SU}) we have  
\begin{align}\label{prod}
\Delta S_U=\Delta S_S+\frac{\Delta E_R}{T}.
\end{align}
Now we can separate the Hamiltonian (\ref{HU}) in three parts $H=H_S+H_I+H_R$. As the universe is a closed system, $\Delta E_R=-(\Delta E_S+\Delta E_I)$. The system energy can be easily calculated using the reduced density operator, however, calculating the interaction energy is not so trivial. To evaluate the latter we will make use of the Hamiltonian of Mean Force (HMF) \cite{HMF1,HMF2,HMF3}. Using the HMF, $H^*$, the interaction energy in equilibrium is given by $E_I=\langle H^*\rangle-E_S$ \cite{HMF2}, and, as the interaction energy at $t=0$ is zero, the entropy production is given by
\begin{align} \label{SU 1}
    \Delta S_U=\Delta S_S-\frac{\langle H^* \rangle}{T}+\frac{\hbar \omega_0}{2T}.
\end{align}
In what follows we will define the HMF and obtain its analytical expression. We will also be able to extend the definition of the HMF for systems which have not yet reached equilibrium or, in other words, are time dependent.

Finally, we have reached the considerations about the central issue of this letter. In order to implement our hypothesis, it would be desirable to write the very general reduced density operator in a form useful for all the applications we want to address. A particularly interesting form is that of an exponential of an operator  which depends only on the observables ($q$ and $p$) of the system of interest. Since this  resembles the Gibbs operator itself, it allows us to make all the necessary extensions to the more general situation in a straightforward way. Working on obtaining that specific form of $\rho_S$ is what we shall do from now onwards.

The time evolution of the reduced density operator of the system described by (\ref{HU}) in the coordinate representation is \cite{caldeira1,caldeira2,caldeira3,Weiss,Breuer}
\begin{align} \label{reduced density f}
\rho_S( x,  y ,t) = \int\!\!\!\int dx' dy' \mathcal{J}( x, y,t; x',  y',0)\rho_S( x',  y',0),
\end{align}
where $\mathcal{J}( x, y,t; x',  y',0)$ is the superpropagator of the system which determines its time evolution from a given initial state. The exact form of the superpropagator which is determined by path integral techniques is left for the supplementary material.

 Performing the path integrals to find the superpropagator is a challenge, and, in this paper, we will analyze one of the few cases in which an analytic solution is available. We will address a system coupled with an ohmic heat bath (discussed in the supplementary material) together with $V(q)=m\omega_0^2 q^2/2$. Now, without loss of generality, we impose that our system is initially in  its {\it{non-interacting}} ground state. Using the coordinates of the ``center of mass" and ``relative position", respectively $q=(x+y)/2$ and $\xi=x-y$, of its wave packet, we can write the initial density operator of our system, $\langle x' | 0 \rangle \langle 0 | y' \rangle$, as
\begin{align} \label{ground}
\rho_S(q',\xi',0)=\frac{1}{\sqrt{2 \pi \sigma^2}}\text{exp}{-\bigg\{\frac{q'^2}{2 \sigma^2}+\frac{\xi'^2}{8 \sigma^2}\bigg\}}.
\end{align}
where $\sigma=\sqrt{\hbar/2 m \omega_0^2}$.

As we are dealing with a fully quadratic system (see supplementary material), we can solve the path integral exactly, and obtain $\mathcal{J}( q,\xi,t; q', {\xi}',0)$ which describes the time evolution of any damped quantum harmonic oscillator \cite{caldeira1,caldeira2,caldeira3}. Once we have obtained this superpropagator, we can perform the ordinary double integral in (\ref{reduced density f}), in the new variables $q'$ and ${\xi}'$, to follow the time evolution of the reduced density operator of the system.  

After having performed all these manoeuvres we may write the position representation of the reduced density operator as
\begin{align} \label{rho 1SHO}
\rho_S(q,\xi,t)=\frac{1}{\mathcal{Z}(t)} \text{exp}-\bigg\{\frac{q^2}{2\sigma^2(t)}+F(t)\xi^2-i D(t) q\xi \bigg\},
\end{align}
where its normalization function is given by $\mathcal{Z}(t)=\sqrt{2 \pi \sigma^2(t)}$. The functions $\sigma^2(t)$, $F(t)$ and $D(t)$ are time dependent functions which encompass thermal and dissipative effects of the particle motion. Their exact forms can be found in the supplementary material.

Our main goal in this paper is to study the entropy, and the entropy production, of this system for arbitrary temperatures and couplings. As the von Neumann entropy is given by (\ref{von-Neumann}), in order to evaluate it, we need to solve the integral
\begin{align}
S=-k_B \int\!\!\!\int dx dy \langle x |\rho_S| y \rangle \langle y | \ln \rho_S |x \rangle.
\end{align}
However, evaluating  such an expression  is not a trivial matter if we have only the position representation of the reduced density operator. Note that $\langle y | \ln \rho_S |x \rangle= \ln \ \rho_S (x,y)$ if, and only if, $\rho_S(x,y)$ is diagonal in the position representation which is not true in general, and particularly not true for our system. Therefore, to obtain the entropy of the system of interest in the full quantum description, we must either find a way to obtain the position representation of the logarithm of its reduced density operator or circumvent this problem by using another technique. Firstly, however, let us  the briefly revisit the evaluation of the entropy  of the system in the classical limit.

In the limit $k_B T \gg \hbar \omega_0$, the Wigner transform of the density operator of a particle, defined as
\begin{align}
\rho_W(q,p,t) =\int d \xi \frac{e^{-i \frac{p\xi}{\hbar}}}{2 \pi \hbar} \bigg \langle q+\frac{\xi}{2} \bigg|\rho_S \bigg|q-\frac{\xi}{2} \bigg\rangle,
\end{align}
connects smoothly with the density of points in its classical phase space. Henceforth, we will omit the time dependence of functions $\sigma(t)$, $F(t)$ and $D(t)$ in (\ref{rho 1SHO}). Performing the Wigner transform of the function (\ref{rho 1SHO}) we obtain
\begin{align} \label{reduced W}
\begin{aligned}
\rho_W(q,p,t)=\frac{1}{2\pi \hbar W} &\exp{-\bigg\{\frac{1}{4\hbar^2 F}p^2} \\
&+\frac{D}{2 \hbar F}pq+\frac{D^2\sigma^2+ 2F}{4F\sigma^2}q^2 \bigg\}
\end{aligned}
\end{align}
where we have defined the function $W(t)=\sqrt{2F(t)\sigma^2(t)}$ for convenience since the entropy, both in the classical and  quantum regimes, are described in terms of this function.

For classical systems with well-defined distribution functions there is a clear prescription \cite{Landau,Salinas,LeBellac} for obtaining the entropy provided by the Boltzmann $H$-theorem which states that

\begin{align}
S_{W}&=- k_B \int\!\!\!\int dp dq \ \rho_W(q,p,t) \ln 2\pi \hbar \rho_W(q,p,t).
\end{align}
Using (\ref{reduced W}) we obtain
\begin{align} \label{S W}
S_W=k_B (1 +\ln W).
\end{align}
Notice that this expression, and consequently its use in (\ref{prod}) to obtain the entropy production, is only valid for high temperatures, when quantum effects are not present. So, if we wish to obtain a valid expression for the entropy which takes into account quantum effects we must take a completely different approach.

Let us propose an ansatz for the reduced density operator of the system. Suppose it can be written as
\begin{align} \label{ansatz}
\hat{\rho}_S =\frac{1}{Z_{a}} \text{exp}-[a(t)q^2+b(t)p^2+c(t)(qp+pq)]
\end{align}
where $a(t)$, $b(t)$, and $c(t)$ are functions of time, whereas $q$ and $p$ are now operators. We will impose that the coordinate representation of $\hat{\rho}_S$ in (\ref{ansatz}) is given by the function (\ref{rho 1SHO}), or
\begin{align} \label{condition}
\langle x | \hat{\rho}_S (t) |y \rangle = \rho_S (q,\xi,t).
\end{align}
Notice that using this ansatz, a simple expression for the entropy of the system results;
\begin{align} \label{S A}
\begin{aligned}
S=k_B[a\langle q^2 \rangle + b \langle p^2 \rangle + c \langle \{ q,p\} \rangle+\ln Z_{a}].
\end{aligned}
\end{align}

We now have to find expressions for the functions $a(t)$, $b(t)$, $c(t)$ and $Z_{a}(t)$ which will satisfy the condition (\ref{condition}). To find such expressions we will implement a path integral representation for (\ref{ansatz}) in terms of an  auxiliary parameter $u$ which will play the same role as $\beta$ in Euclidean path integrals or time in the propagator. However, our parameter $u$ will be a purely mathematical tool with no physical meaning, and, in the end, we will set it equal to unity. Notice that this procedure is quite general and in no way limited to our ansatz.

Stating it more clearly, we will perform a path integral in $u$ to find $\langle x|e^{-A(q,p)u}|y \rangle |_{u=1}$. Although this is a novel idea and extends the usefulness of path integral techniques to situations where the time dependence (or $\beta$ dependence) is not as trivial as the standard cases, the formalism from now own is straightforward and thus left to the supplementary material. The position representation of the operator $\langle x|e^{-A(q,p)u}|y \rangle$ for the most general $A(q,p)$ is given by
\begin{align} \label{path int}
\int\!\!\!\int\limits_y^x   [d q(u')  d p(u')] \exp{\int\limits_0^1 du' \bigg\{\frac{i}{\hbar}p \dot q-A(q,p)\bigg\}}.
\end{align}
Evaluating this path integral for our ansatz (\ref{ansatz}), we may write the position representation of the reduced density operator as
\begin{align}
\begin{aligned}
\langle x |\hat{\rho}_S| y \rangle =\frac{G(t)}{Z_{a}} &\exp{-\alpha \big\{ x^2[\Omega \coth \Omega + \Gamma]}\\
&-2xy \frac{\Omega}{\sinh \Omega}+y^2[\Omega \coth \Omega - \Gamma]\big\}
\end{aligned}
\end{align}
where $\Gamma=i 2 \hbar c$, $\Omega=2\hbar \sqrt{ab-c^2}$,
\begin{align} \label{G(t)}
G(t)=\oint \mathcal D y(u) \exp{-\int_0^1 du'\big\{\alpha \dot y^2+\psi \dot y y +\phi y^2 \big\}},
\end{align}
$\alpha=1/4 \hbar^2 b$, $\psi= ic/\hbar b$ and $\phi=a-c^2/b$. Notice that, as the function $G(t)$ depends solely on time, it is significant only as a normalization. We may also use the definition of $q$ and $\xi$ in terms of $x$ and $y$ to write 
\begin{align}
\begin{aligned}
\rho_S(q,\xi,t)=\frac{1}{\mathcal Z} &\exp{-\bigg\{x^2\bigg[\frac{1}{8\sigma^2}+F+i\frac{D}{2}\bigg]} \\
&-2xy\bigg[F-\frac{1}{8\sigma^2}\bigg]+y^2 \bigg[\frac{1}{8\sigma^2}+L-i\frac{D}{2}\bigg]\bigg\}.
\end{aligned}
\end{align}
Then, imposing $\langle x | \hat{\rho}_S (t) |y \rangle = \rho_S (q,\xi,t)$ we find the system of equations
\begin{align}
\left\{
\begin{aligned}\label{KLM}
&Z_{a} = G(t) \mathcal Z \\
&\alpha \Omega \coth \Omega = \frac{1}{8\sigma^2}+F \\
&\frac{\alpha \Omega}{\sinh \Omega}=F-\frac{1}{8\sigma^2} \\
&\alpha \Gamma = i \frac{D}{2}
\end{aligned} \right.
\end{align}
The first equation is completely independent of the others and to solve it we will have to explicitly evaluate the functional integral (\ref{G(t)}). The other three, however, form a system of coupled transcendental equations. Although the latter, in general, are not solvable in terms of ordinary functions, this system is one of the few exceptions where an analytic solution can be found. To find such a solution notice that, in order to perform the saddle point approximation, and integrate out the momentum in (\ref{path int}), we implicitly assumed that $\alpha\Omega \neq 0$. Therefore, we may divide the second and third equations above by this product, and use a logarithmic expression for the inverse of the hyperbolic cosine to obtain
\begin{align}
\Omega = \ln{\frac{2W+1}{2W-1}}.
\end{align}

Using the definitions of $\Omega$, $\Gamma$, and $\alpha$ in terms of $a$, $b$, and $c$ we may express them in terms of the well know functions $D$, $F$ and $\sigma^2$
\begin{align} \label{abc}
\left\{
\begin{aligned}
&a =\frac{D^2\sigma^2+ 2F}{4F\sigma^2} W \Omega \\
&b =\frac{1}{4 \hbar^2 F}  W \Omega \\
&c=\frac{D}{4 \hbar F}  W \Omega
\end{aligned} \right.
\end{align}
We would like to remark that these three functions differ from the ones accompanying the Wigner transformation of the reduced density operator (\ref{reduced W}) by the term $W \Omega$ which emphasizes the conceptual difference between the Wigner representation of the reduced density operator and the operator itself.

To obtain the partition function $Z_{a}(t)$ we need  $G(t)$ which results from the evaluation of (\ref{G(t)}). The procedure to evaluate the latter is the standard one used in the solution of quadratic path integrals, and, therefore, it is left for the supplementary material. Having a closed form for $G(t)$ we may obtain the partition function of the system
\begin{align} \label{Z(t)}
Z_{a}(t)=\frac{\sqrt{4W^2-1}}{2} .
\end{align} 

\begin{figure}
\centering
\includegraphics[scale=0.30]{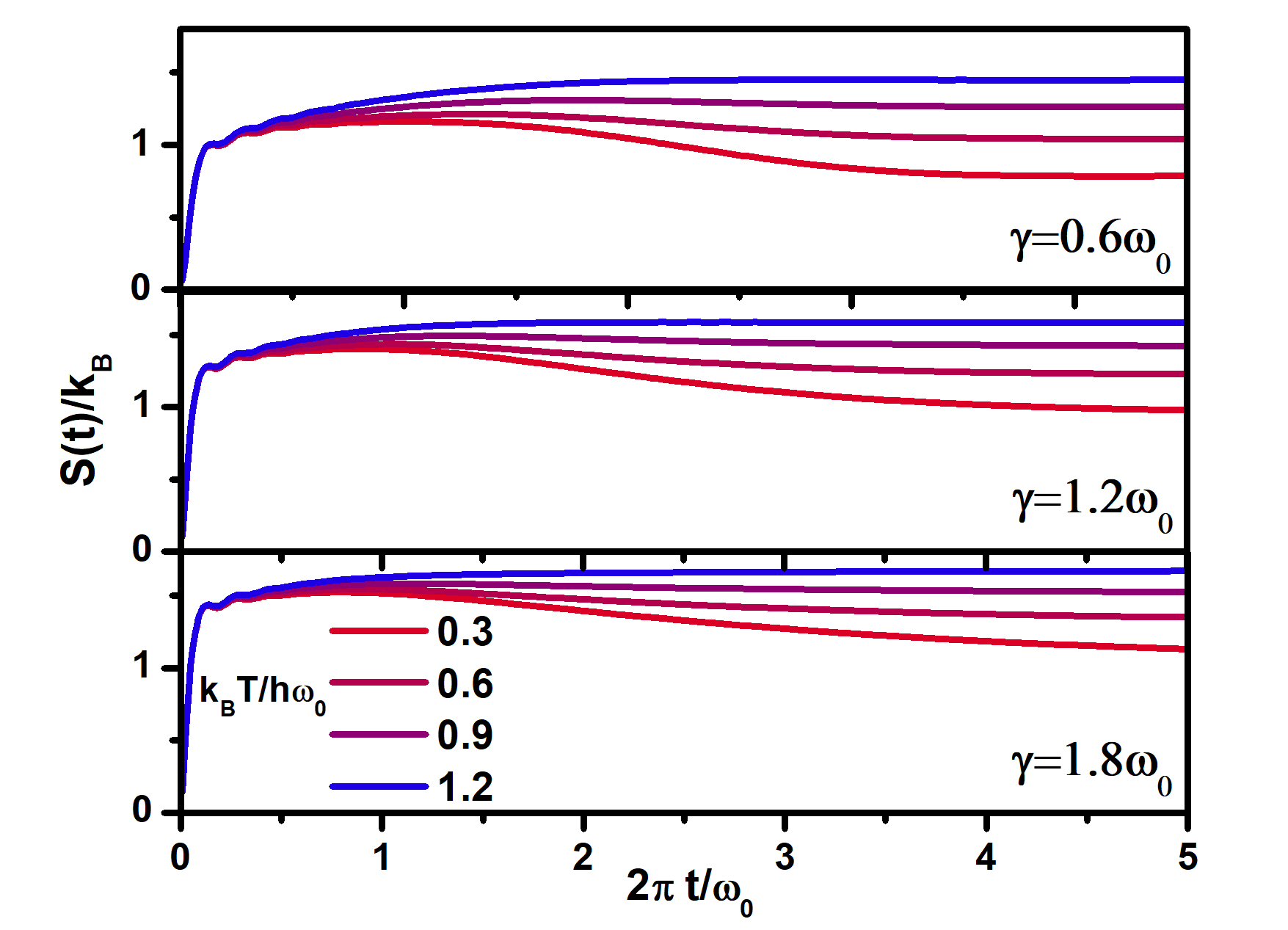}
\caption{The entropy as a function of time given by equation (\ref{S Q}) for different temperatures and coupling strengths. Notice that when the temperature is low and quantum effects are prominent the entropy starts at zero, goes to a maximum value, and then decreases until it reaches its equilibrium value. That is, the entropy is not a monotonic function of time for low T.}
\label{Fig1}
\end{figure}

Now using (\ref{rho 1SHO}) to calculate $\langle q^2 \rangle$, $\langle p^2 \rangle$ and $\langle \{ q, p\} \rangle$ we obtain
\begin{align} \label{x, p, xp}
\left\{
\begin{aligned}
&\langle q^2 \rangle = \sigma^2 \\
&\langle p^2 \rangle = \hbar^2(2F+D^2\sigma^2) \\
&\langle \{ q, p\} \rangle = -\hbar2 D\sigma^2.
\end{aligned} \right.
\end{align}
Finally, substituting (\ref{abc}), (\ref{Z(t)}), and (\ref{x, p, xp}) in (\ref{S A}), we obtain an expression for the entropy of a harmonic oscillator coupled to a heat bath valid at any temperature and coupling constant,
\begin{align} \label{S Q}
\begin{aligned}
S= k_B\left[ W \ln{\frac{2W+1}{2W-1}}+\ln \sqrt{4W^2-1}-\ln 2 \right].
\end{aligned}
\end{align}
The only restriction to the validity of our expression is that the heat bath must be ohmic. Given that, which is a quite general case, this expression has a wide range of applicability. The time evolution of the entropy can be seen in (Figure \ref{Fig1}). Also, using (\ref{x, p, xp}) we can express $W(t)$ in terms of the average of the position and momentum operators as
 \begin{align} \label{W}
W(t)=\frac{1}{\hbar}\sqrt{\langle p(t)^2\rangle\langle q(t)^2 \rangle - \langle \{ p(t),q(t) \} \rangle /4}.
\end{align}
Using this expression, we may obtain $W(t)$ in the Heisenberg picture or use (\ref{p2 and q2}) to obtain the equilibrium value of $W(t)$.

Now that we have the final form of the entropy we can study the entropy production using (\ref{SU 1}). The last ingredient missing is an explicit form for the average of the HMF. The latter is defined by demanding that
\begin{align}
e^{-\beta H^*}=Z_R^{-1}\text{tr}_R[e^{-\beta H}],
\end{align}
meaning that, the HMF is an effective operator, with which we can build a Gibbs like operator, which coincides with the reduced density operator of the system in equilibrium. But this is just one especial case of our ansatz (\ref{ansatz}), namely the case where we take the limit of time going to infinity. Then, the HMF is given by
\begin{align} \label{H*}
    H^*=\frac{1}{\beta}\lim_{t \rightarrow \infty}[aq^2+bp^2+c(qp+pq)].
\end{align}
In other words, our ansatz (\ref{ansatz}) can be seen as a definition of the HMF out of equilibrium. 

To obtain the entropy production we need $\langle H^*\rangle$. Therefore, taking the average of (\ref{H*}) using (\ref{abc}), (\ref{x, p, xp}), and (\ref{W}), we obtain an analytical expression for the average of the HMF as
\begin{align} \label{HMF F}
\langle H^* \rangle= k_BT \ W \ln{\frac{2W+1}{2W-1}}.
\end{align}
Next, using (\ref{SU 1}) we obtain the entropy production
\begin{align} \label{Su F}
\Delta S_U = k_B\left[\ln\sqrt{4W^2-1}-\ln 2\right]+\frac{\hbar \omega_0}{2T}.
\end{align}
Expressions (\ref{ansatz}), (\ref{abc}), (\ref{S Q}), (\ref{HMF F}), and (\ref{Su F}) are the main results of our paper. Those, provide us with; (i) an analytical expression for the reduced density operator itself, and not just its coordinate representation, (ii) the entropy related to the system, (iii) its HMF, and (iv) the entropy production. Moreover, we have furnished the full time dependence of all those expressions which are valid for any temperature and coupling strength.

Now, let us study some limits of our expressions. Our system was in the ground state at the initial instant, which is a pure state, and thus have zero entropy. We can use the analytic expression (\ref{W}) for $W(t)$ to find $\lim_{t \rightarrow 0}W(t)=1/2$ and then $\lim_{t\rightarrow 0}S(t)=0$. Now, in equilibrium ($t \rightarrow \infty$), we find that $\langle q^2 \rangle$ and $\langle p^2 \rangle$ are given by (\ref{p2 and q2}) and that $\langle\{q,p \} \rangle=0$.  

For the weak coupling regime ($\gamma \rightarrow 0$) the entropy of our system is 
\begin{align} \label{S gamma=0}
\begin{aligned}
S^{eq}_{\gamma=0}=\frac{\hbar \omega_0}{2 T}\coth{\frac{\hbar \omega_0}{2 k_B T}}-k_B\ln{\sinh{\frac{\hbar \omega_0}{2 k_B T}}}-k_B\ln 2,
\end{aligned}
\end{align}
which is exactly the expression for the entropy of a quantum harmonic oscillator in the canonical ensemble given by standard statistical physics \cite{Landau,Salinas}. The average of the HMF also gives us just the energy of a quantum harmonic oscillator in the canonical ensemble
\begin{align} \label{HMF g=0}
    \langle H^* \rangle_{\gamma=0}=\frac{\hbar \omega_0}{2} \coth{\frac{\hbar \omega_0}{2k_BT}}.
\end{align}
Finally the entropy production is given by
\begin{align} \label{SU g=0}
    \Delta S_{U , \gamma=0}=\frac{\hbar \omega_0}{2T}-k_B\ln{\sinh{\frac{\hbar \omega_0}{2 k_B T}}}-k_B\ln 2,
\end{align}
\begin{figure}
\centering
\includegraphics[scale=0.30]{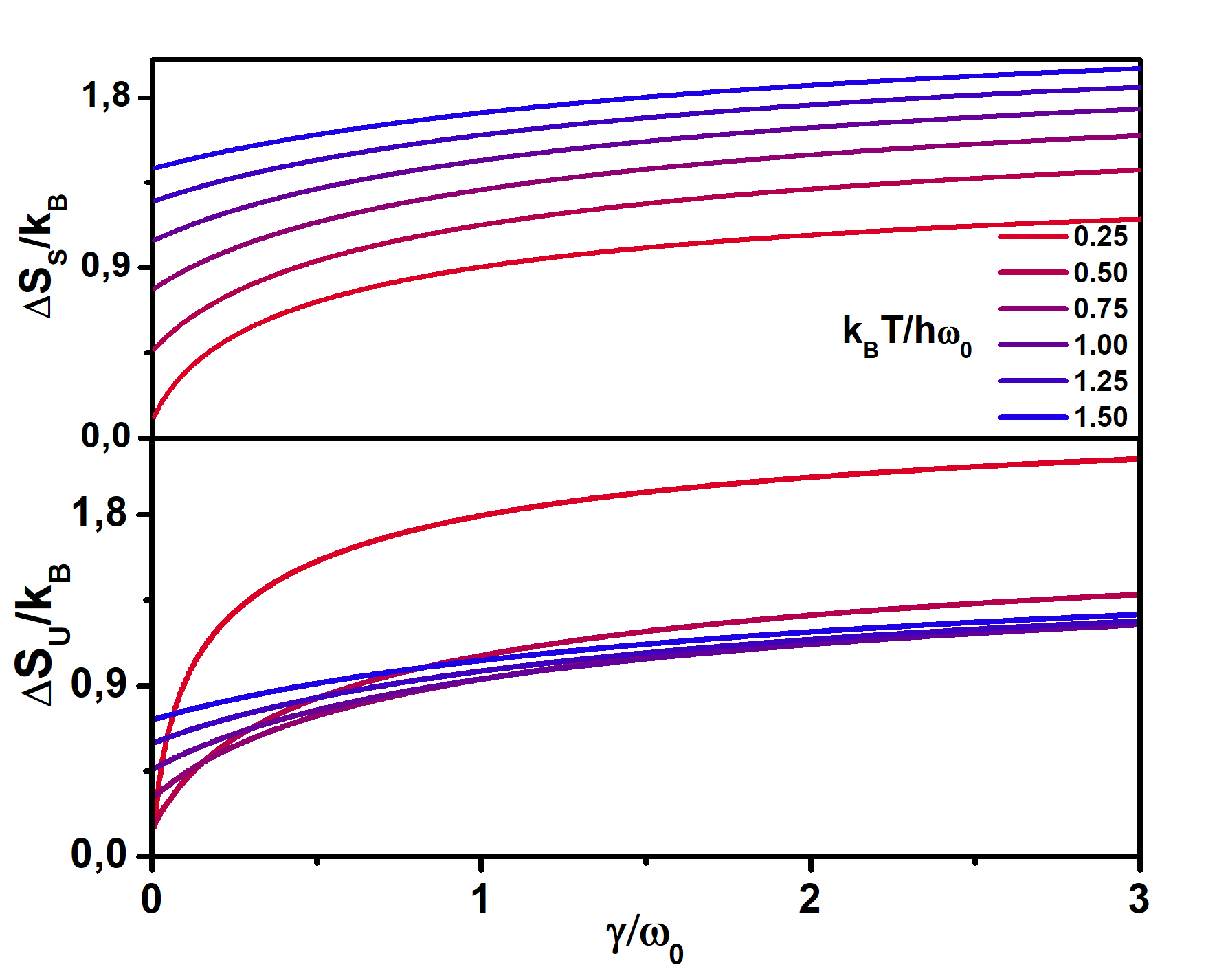}
\caption{$\Delta S_S$ and $\Delta S_U$ as a function of $\gamma$ for different values of $T$. Notice that the entropy production is always positive for any temperature and coupling strenght.}
\label{Fig2}
\end{figure}

\noindent which is always positive obeying the second law of thermodynamics. As the final values of thermodynamical quantities are independent of the initial state, expression (\ref{S gamma=0}) is also valid for any initial state of our system, and not only the ground state (\ref{ground}). The equilibrium value of the energy (\ref{HMF g=0}) is also unaffected by the choice of the initial state. Therefore, for eigenstates of the simple harmonic oscillator $\Delta S>0$ and $\Delta S_U>0$, in the weak coupling regime.

In the high temperature limit, the equipartition theorem holds and the entropy is given by
\begin{align} \label{S highT}
    S^{eq}_{\gamma=0, k_BT\gg\hbar \omega_0}=k_B\left(1+\ln \frac{k_BT}{\hbar \omega_0} \right)
\end{align}
which is also the weak-coupling regime of (\ref{S W}), the entropy obtained using the Boltzmann  $H$-theorem applied to the Wigner transform of the reduced density operator. The entropy production in this regime is just $\Delta S_U=k_B \ln (k_B T/\hbar \omega_0)$, which is also positive.

Let us now explore the strong coupling regime ($\gamma \gg \omega_0$) in equilibrium. The procedure for studying this regime is not so straightforward as for the former cases. However we are able to solve the integrals (\ref{p2 and q2}) for $k_BT\gg \hbar \omega_0$ or $k_BT\ll \hbar \omega_0$ using partial fraction decomposition. 

In the high temperature limit, when thermal effects dominates over all the other energy scales of the problem, the result is the same as $\gamma\ll\omega_0$, meaning that the equipartition theorem still holds and the entropy is given by (\ref{S highT}), which also implies that $\Delta S_U>0$. Nevertheless, in the low temperature regime, things are quite different. We can show that 
\begin{align}
\begin{aligned} \label{gamma grande}
    \langle q^2\rangle^{eq}_{ \gamma\gg\omega_0}
    =\frac{2\hbar}{m\pi\gamma}\ln \frac{2\gamma}{\omega_0},&&
    \langle p^2\rangle^{eq}_{  \gamma\gg\omega_0}=\frac{2m\hbar\gamma}{\pi}\ln \frac{\gamma\Omega}{2\omega_0^2}.
\end{aligned}
\end{align}
Using these results in (\ref{W}), and the resulting expression in (\ref{S Q}) and (\ref{Su F}), we can find the entropy of the system and the entropy production, which are both always positive.

In this letter we have developed a path integral technique with respect to a fictitious  parameter which enabled us to get the reduced density operator itself, in terms of the observables of the system, instead of only its representation in a given basis. Path integrals are a valuable tool for finding the representations of operators which have a free parameter like time (for the time evolution operator) or inverse temperature (for Euclidean path integrals). Our technique extends the usefulness of path integrals to more general operators whose time and temperature dependence are not as trivial as those standard cases.

This technique proved to be valuable in the study of open quantum systems. Using the reduced density operator we have obtained from our path integral approach - equations (\ref{ansatz}) and (\ref{abc}) - we have evaluated analytical expressions for the entropy, entropy production, and the HMF of a harmonic oscillator coupled with a heat bath (\ref{S Q}), (\ref{Su F}), and (\ref{HMF F}). In the limits of weak coupling and high temperatures these expressions give us the well-known results from statistical mechanics (\ref{S gamma=0}-\ref{S highT}). Those expressions also allowed us to study the problem in situations to which the standard approach of statistical mechanics cannot be applied.

We have shown analytically that, in our specific example, the entropy of the system, and of the universe always increase irrespective of temperature or coupling strength (Figure \ref{Fig2}). That is, the second law of thermodynamics holds true for any temperature or coupling strength. In some previous works it was shown that for very weakly coupled open quantum systems, when one completely ignores the thermal reservoir, the entropy production is positive \cite{Spohn,RevTerm,Deffner,Deffner2,Oliveira,Aurell} and others even argued about a violation of the second law of thermodynamics when strong coupling is considered \cite{Theo}. Our approach extend those early results for strong coupling regimes showing that the entropy of the whole universe always increases.

The success of our approach stems form the use of (\ref{prod}) to calculate the entropy production. Most approaches in the literature use simply
\begin{align} \label{SU FS}
    \Delta S_U=-\Delta F_S/T
\end{align}
(note that expressions using relative entropy are a consequence of this expression). This expression is obtained from standard statistical mechanics in the weak coupling regime and its validity is assumed to still hold for strong couplings. However, using this latter expression we obtain a negative entropy production in the strong coupling regime. We argue here that this is not a violation of the second law, it is just a misinterpretation of the validity of (\ref{SU FS}). By using this expression we ignore the interaction energy that comes from the coupling which led us from (\ref{prod}) to (\ref{SU 1}) by the use of the HMF. One important thing to notice is that, if we agreed that the energy of the system must be given by the average of the HMF, and not only of the system Hamiltonian, and used (\ref{SU FS}), we would still reach expression (\ref{SU 1}) and therefore have a positive entropy production. We believe however that the interpretation we provided is more accurate since it comes from the use of (\ref{SU}), the additivity of the entropy, and a simple consideration of the heat exchanged between the reservoir and the system.

One final note is that, as we all know, the second law of thermodynamics only precludes the entropy of the whole universe to decrease, and not necessarily of the system only. We can see for example in (Figure \ref{Fig1}) that the time evolution of the entropy is not monotonic, but as the entropy of the universe still increases this is not a violation of the second law. Furthermore, in our study the system was in a pure state at time $t=0$ and evolved to equilibrium. As the entropy of a pure state is zero, then, by preparation, $\Delta S_S>0$. However, we could have prepared our system in a mixed state and then connected it with a reservoir at a temperature lower than the average initial energy of the system, in such a way that the equilibrium state has less entropy than the the initial state. In this case $\Delta S_S<0$ but still $\Delta S_U>0$ in accordance with the second law of thermodynamics.

Financial support for this work was provided by the Brazilian funding agencies, CAPES and CNPq, under the projects numbers 140553/2018-5 and 302420/2015-0, respectively.


\onecolumngrid

\pagebreak

\begin{center}
  \textbf{\large von Neumann entropy and the entropy production of a damped harmonic oscillator\\Supplementary Material}\\[.2cm]
 G. A. Weiderpass$^{1,*}$ and A. O. Caldeira$^{1}$\\[.1cm]
  {\itshape ${}^1$Instituto de  F\'\i sica Gleb Wataghin, Universidade Estadual de Campinas, 13083-859, Campinas, SP, Brazil}\\
\end{center}

\setcounter{equation}{0}
\setcounter{figure}{0}
\setcounter{table}{0}
\setcounter{page}{1}
\renewcommand{\theequation}{S\arabic{equation}}
\renewcommand{\thefigure}{S\arabic{figure}}
\renewcommand{\bibnumfmt}[1]{[S#1]}
\renewcommand{\citenumfont}[1]{S#1}

\section{Spectral Density for Brownian Motion}

The model for quantum Brownian motion, which is the paradigm of open quantum systems in the scientific literature, \cite{caldeira3-S,Weiss-S,Breuer-S} is described by the Hamiltonian
\begin{align} \label{HU-S}
\begin{aligned}
H = \frac{p^2}{2 m} + V(q)+\sum_{j}^{} \Big[ \frac{p_{j}^2}{2 m_{j}} + \frac{m_{j} \omega_{j}^2}{2} \Big(q_{j} -\frac{C_{j}}{m_{j} \omega_{j}^2 }q \Big)^2\Big],
\end{aligned}
\end{align}
with the prescription that the spectral function is given by
\begin{align}\label{spectral-S}
J(\omega)\equiv\frac{\pi}{2} \sum_j^{} \frac{C_{j}^2}{m_{j} \omega_{j}} \delta(\omega-\omega_{j}).
\end{align}
If the system is in contact with an ohmic heat bath, which is the most ubiquitous of cases, then the spectral function should be modelled by
\begin{eqnarray}
J(\omega)=\left\{\begin{array}{ccc}
  \eta \, \omega & \textrm{if} & \omega<\omega_c \\
  0 & \textrm{if} & \omega>\omega_c. \\
\end{array} \right.
\label{ohmicJ}
\end{eqnarray}
which, in the classical limit, and for times $t>>\omega_c^{-1},$ reproduces the Langevin equation
\begin{eqnarray}
m \, \ddot{q} + \eta\, \dot{q} + V^{\prime}(q) = f(t) ,\label{langevin}
\end{eqnarray}
with
\begin{align}
\left\langle f(t) \right\rangle = 0 & & \text{and} & & \left\langle f(t) \, f(t^{\prime}) \right\rangle = 2 \, \eta \,
k_{B} \, T \, \delta \left( t - t^{\prime} \right) . \label{white}
\end{align}
Here, $\eta$ is the damping constant which is related to the relaxation frequency, $\gamma$, through $\eta=2m\gamma$. In this model, the reservoir is held at constant temperature $T$. 

\section{Reduced Density Operator}

The time evolution of the reduced density operator of a system described by (\ref{HU-S}, \ref{spectral-S}) in the coordinate representation is given by
\begin{align} \label{reduceddensityf-S}
\rho_S( x,  y ,t) = \int dx' dy' \mathcal{J}( x, y,t; x',  y',0)\rho_S( x',  y',0).
\end{align}
where
\begin{align} \label{propagator}
\begin{aligned}
\mathcal{J}( x, y,t; x',  y',0)=\int\limits_{ x'}^{x} \int\limits_{y'}^{y} \mathcal D  x(t) \mathcal D  y(t)
\exp \left\{\frac{i}{\hbar}\{S_S[ x(t)]-S_S[ y(t)]\} -\frac{1}{\hbar}\phi[x(t),y(t)]\right\},
\end{aligned}
\end{align}
is the so-called superpropagator of $\rho_S(t)$, with $S_S[x(t)]$ being the classical action of the system 
\begin{align}
&S_S[ x(t)]= \int\limits_0^t dt' \left\{ \frac{m}{2}\dot x^2 - V( x) \right\},
\end{align}
and $\phi[x(t),y(t)]$, which is the Feynman-Vernon influence functional, given by
\begin{align} \label{influence functional-S}
\begin{aligned}
&\phi[x(t),y(t)]=\int\limits_0^t \int\limits_0^{t'} dt' dt'' \{x(t')-y(t') \}\{ \mathcal{L}(t'-t'') x(t'')-\mathcal{L}^*(t'-t'') y(t'') \}+ i \sum\limits_k \frac{C_k^2 }{2m_k \omega_k^2} \int\limits_0^t dt' \{x^2(t')-y^2(t') \},
\end{aligned}
\end{align}
where
\begin{align}
&\mathcal{L}(t)= \frac{1}{\pi} \int\limits_0^\infty d \omega J(\omega) \Big( \coth\frac{\hbar \omega}{2 k T}\cos \omega t-i \sin \omega t \Big), 
\end{align}
and
\begin{align}
\sum_k \frac{C_{k}^2}{2 m_{k} \omega_{k}^2}= \frac{1}{\pi} \int\limits_0^{\omega_c} d\omega \frac{J(\omega)}{\omega}.
\end{align}
It should be stressed that the last term on the r.h.s. of (\ref{influence functional-S}) is the well-known counter-term already exhaustively discussed in the literature \cite{caldeira3-S,Weiss-S,Breuer-S}.

For an ohmic heat bath and $V(q)=m\omega_0^2 q^2/2$, we can solve the double functional integral given by equation (\ref{propagator}), rewritten in terms of the new paths, $q(t')=(x(t')+y(t'))/2$ and $\xi(t')=x(t')-y(t')$, to obtain the superpropagator $\mathcal{J}( q,\xi,t; q', {\xi}',0)$ which describes the time evolution of any damped quantum harmonic oscillator. Since we are dealing with a fully quadratic system, that functional integral can be exactly evaluated \cite{caldeira3-S}. Once we have obtained this superpropagator, we can perform the ordinary double integral in (\ref{reduceddensityf-S}), in the new variables $q'$ and ${\xi}'$,  to follow the time evolution of the reduced density operator of the system.  

After having performed all these manoeuvres we obtain the position representation of the reduced density operator as
\begin{align} \label{rho 1SHO-S}
\rho_S(q,\xi,t)=\frac{1}{\mathcal{Z}(t)} \text{exp}\bigg[-\frac{q^2}{2\sigma^2(t)}-F(t)\xi^2+i D(t) q\xi \bigg],
\end{align}

where the functions $\sigma^2(t)$, $F(t)$ and $D(t)$ are given  by
\begin{align} \label{KLM-S}
&\begin{aligned}
\sigma^2(t)=\frac{\sigma^2 K_1^2(t)+2\hbar C_1^2(t)}{N^2(t)}
\end{aligned}\\
&\begin{aligned}
F(t)=\frac{A(t)}{\hbar}+\frac{\sigma^2L(t)}{2\hbar^2}-\frac{[\sigma^2 K_1(t)L(t)-\hbar B(t)]^2}{2\hbar^2 \sigma^2(t) N^2(t)}
\end{aligned} \\
&D(t)=K_2(t)-\frac{\sigma^2 K_1(t)L(t)-\hbar B(t)}{ \sigma^2(t) N(t)}
\end{align}
with
\begin{align}
C_1(t)=C(t)+\frac{\hbar}{8\sigma^2}, & & & K_1(t)=K(t)+m\gamma, & & K_2(t)=K(t)-m\gamma, \\
K(t)=m \omega \cot \omega t, & & & L(t)=\frac{m \omega e^{-\gamma t}}{\sin \omega t}, & & M(t)=\frac{m \omega e^{\gamma t}}{\sin \omega t}.
\end{align}
where $\omega=\sqrt{\omega_0^2 - \gamma^2}$. Lastly, these expressions depend on a set of three function $A(t)$, $B(t)$ and $C(t)$ of the form
\begin{align}
f(t)= \frac{m \gamma}{\pi} \int_0^\Omega d \nu \nu \coth \frac{\hbar \nu}{2 k_B T} f_\nu(t),
\end{align}
where their generating functions, $A_\nu (t)$, $B_\nu (t)$ and $C_\nu (t)$, are given by the following integral forms 
\begin{align}
\begin{aligned}
A_\nu(t)= \frac{e^{-2\gamma t}}{\sin ^2 \omega t}& \int_0^t \int_0^t dt' dt'' e^{\gamma(t'+t'')}
 \cos \nu (t'-t'') \sin \omega t' \sin \omega t'' \\
B_\nu(t)= \frac{e^{-\gamma t}}{\sin ^2 \omega t}& \int_0^t \int_0^t dt' dt'' e^{\gamma(t'+t'')}
\cos \nu (t'-t'') \sin \omega t' \sin \omega(t-t'') \\
C_\nu(t)= \frac{1}{\sin ^2 \omega t} &\int_0^t \int_0^t dt' dt'' e^{\gamma(t'+t'')}
\cos \nu (t'-t'') \sin \omega(t-t') \sin \omega(t-t'').
\end{aligned}
\end{align}

\section{Path Integral in the effective parameter}

We will find the position representation of operators of the form $e^{-A(q,p)u}$ by discretizing the parameter $u$ in the following way
\begin{align} \label{prop-S}
\begin{aligned}
&\langle x | e^{-A(q,p) u}|y \rangle =\int\limits_{-\infty}^{\infty}...\int\limits_{-\infty}^{\infty} dx_1...dx_{N-1}
&\langle x | e^{-A(x_{N-1},p_{N-1}) \epsilon}
|x_{N-1}\rangle \langle x_{N-1} |...|x_1 \rangle \langle x_1 |e^{-A(x_{1},p_{1}) \epsilon}|y \rangle
\end{aligned}
\end{align}
where $u=N\epsilon$. As $\epsilon$ is small we may expand the exponentials in series and retain only the first order terms. Then, for each term of the form $\langle x_k|1-A(q,p) \epsilon| x_{k-1} \rangle$, we may insert an identity in the momentum representation and commute the $p$'s and $q$'s in $A(q,p)$ such that all $q$'s are on the left and all the $p$'s on the right. After making the necessary commutations  we may use $\langle x_k|A(q,p) |p_k \rangle=A(x_k,p_k)\langle x_k |p_k \rangle$ and re-exponentiate the expansion. Finally, the matrix element $\langle x_k|e^{-A(q,p) \epsilon}| x_{k-1} \rangle$ can be rewritten as
\begin{align}
\begin{aligned}
&\frac{1}{2 \pi \hbar} \int\limits_{-\infty}^{\infty}dp_k e^{i\frac{p_k}{\hbar}(x_k-x_{k-1})}e^{-\epsilon A(x_k,p_k)}.
\end{aligned}
\end{align}
Taking the limit $N\rightarrow \infty$ with $\epsilon \rightarrow 0$ in a way that $N \epsilon \rightarrow u$, writing derivatives in the parameter $u$ as $\frac{d}{du} x =\dot x$ and making $u \rightarrow 1$ we obtain
\begin{align}
\langle x | e^{-A(q,p)}|y \rangle=\int\!\!\!\int\limits_y^x   [d q(u')  d p(u')] \exp{\int\limits_0^1 du' \bigg\{\frac{i}{\hbar}p \dot q-A(q,p)\bigg\}}.
\end{align}
Notice that until this point the functional measure is just the product of the standard integrals over the $k^{th}$ phase space. However, to evaluate the whole functional integral we first integrate the momentum variables using the saddle point approximation which yields a non-trivial integration measure.

For our specific purposes consider $A(q,p)=aq^2+bp^2+c\{q,p\}$. In this case the saddle point approximation is exact, which enables us to integrate over the momenta leading us to
\begin{align} \label{functional ansatz-S}
\langle x | e^{-A(x,p)}|y \rangle =\int\limits_y^x \mathcal D x(u') \exp{-S[x(u')]},
\end{align}
where the effective parametric action is
\begin{align} \label{acao-S}
S[x(u')]=\int\limits_0^1 du'\{\alpha\dot{x}^2+\psi \dot x x + \phi x^2 \},
\end{align}
with $\alpha=1/4 \hbar^2 b$, $\psi= ic/\hbar b$ and $\phi=a-c^2/b$, and the integration measure is given by
\begin{align}
\int \mathcal D x(u')=\int\limits_{- \infty}^{\infty}...\int\limits_{- \infty}^{\infty} dx_1...dx_{N-1} \Big[ \frac{\alpha}{ \pi  \epsilon } \Big]^{N/2}
\end{align}
At this point one should notice that this parametric functional representation can be performed at any time $t$, which means that, in this process, time can be regarded as a constant. 

\section{Solving the Effective Lagrangian}

We can  write the effective parametric action as an integral of an effective Lagrangian in the parameter $u$ of the form $L=\alpha \dot x^2+\psi \dot x x +\phi x^2$. Then, we find a function which gives an extremum of effective action by solving the Euler-Lagrange equation of this effective Lagrangian. The Euler-Lagrange equation which extremizes the effective parametric action is
\begin{align}
\ddot{\bar{x}} -\Omega^2 \bar x=0.
\end{align}
where $\Omega=2 \hbar \sqrt{ab-c^2}$. Imposing the boundary conditions $\bar x(0)=y$ and $\bar x(1)=x$ we obtain
\begin{align}
\bar x (u)= \frac{1}{\sinh \Omega}\big[x \sinh \Omega u+y\sinh \Omega (1-u) \big].
\end{align}
Now we will perform the following functional variable substitution
\begin{align}
x(u)=y(u)+\bar x(u)
\end{align}
where $y(0)=y(1)=0$. Performing this substitution we obtain
\begin{align}
\langle x | e^{-A(x,p) }|y \rangle=G(t)e^{- S[\bar x, \bar y,t]}
\end{align}
where
\begin{align} \label{G(t)-S}
G(t)=\oint \mathcal D y(u) \exp{-\int_0^1 du'\big\{\alpha \dot y^2+\psi \dot y y +\phi y^2 \big\}}.
\end{align}
Notice that the function $G(t)$ depends exclusively on time, and,  therefore, it is significant only as a normalization. We now substitute $\bar x(u)$ on the effective parametric action and perform the integration in $u'$. Finally, we  write the position representation of the reduced density operator as
\begin{align}
\begin{aligned}
\langle x |\rho_s| y \rangle =\frac{G(t)}{Z} &\exp{-\alpha \Big\{ x^2[\Omega \coth \Omega + \Gamma]}
&-2xy \frac{\Omega}{\sinh \Omega}+y^2[\Omega \coth \Omega - \Gamma]\Big\}
\end{aligned}
\end{align}
where $\Gamma=i 2 \hbar c$.

\section{Obtainign the prefactor G}

To calculate the function $G(t)$ first we integrate by parts the term containing $\psi y \dot y$ on the action and notice that $\dot \psi =0$. We may then write $G(t)$ as the limit of its discretized version
\begin{align}
\begin{aligned}
G(t) = \lim \int dy_1... dy_{N-1} \bigg( \frac{1}{4 \pi\ \hbar^2 \epsilon b} \bigg)^{N/2}
 \exp{- \sum_{K=1}^{N} \Big\{ \alpha \frac{(y_k-y_{k-1})^2}{\epsilon}+ \beta \epsilon y_k^2 \Big\}}.
\end{aligned}
\end{align}
To simplify the notation in the rest of the text "$\lim$" is to be understood as the limit  $N\rightarrow \infty$ with $\epsilon \rightarrow 0$ in a way that $N \epsilon \rightarrow u \rightarrow 1$. Defining $\zeta^T =\begin{pmatrix} y_1 &&...&& y_{N-1}\end{pmatrix}$ and
\begin{align}
\sigma = \frac{\alpha}{\epsilon}
\begin{pmatrix}
\begin{matrix}
 2 &-1 &0 \\ 
 -1& 2& -1 
\end{matrix}&& && 0\\ && \ddots
\\ 0 &&  && 
\begin{matrix}
 2 &-1 \\ 
 -1& 2 
\end{matrix}\end{pmatrix}+ \beta \epsilon 1
\end{align}
we may write
\begin{align}
G(t) = \lim  \bigg( \frac{1}{4 \pi\ \hbar^2 \epsilon b} \bigg)^{N/2} \int d^{N-1} \zeta e^{- \zeta^T \sigma \zeta}.
\end{align}
We may then diagonalize the matrix $\sigma$ and integrate in the new set of variables which makes this matrix diagonal. We obtain
\begin{align}
G(t) = \lim \bigg[ \frac{1}{(4 \pi\ \hbar^2 \epsilon b)^N} \frac{\pi^{N-1}}{\det \sigma} \bigg]^{1/2}.
\end{align}
Now we may separate the whole $u$ dependence in a single function defining
\begin{align}
\begin{aligned}
&\Psi(u)= \lim \epsilon(4 \hbar^2 \epsilon b)^{N-1} \det \sigma
&= \lim \epsilon 
\begin{vmatrix}
\begin{pmatrix}
\begin{matrix}
 2 &-1 &0 \\ 
 -1& 2& -1 
\end{matrix}&& && 0\\ && \ddots
\\ 0 &&  && 
\begin{matrix}
 2 &-1 \\ 
 -1& 2 
\end{matrix}\end{pmatrix}+\Omega^2 \epsilon^2 1
\end{vmatrix}
\end{aligned}.
\end{align}
To perform the limit let us define the determinant of the matrix inside the bars above as $p_{j-1}= (4 \hbar^2 \epsilon b)^{j-1} \det \sigma$ and note that $\Psi(u)=\lim \epsilon p_{N-1}$. We may then expand this determinant in its minors and rearrange the terms obtaining
\begin{align}
\frac{p_{j+1}-2p_j+p_{j+1}}{\epsilon^2}= \Omega^2 p_j.
\end{align}
Performing the adequate limits we see that
\begin{align}
    G(t)=\sqrt{\frac{\alpha}{\Psi(1)}}
\end{align}
where $\Psi(u)$ satisfies the equation
\begin{align}
    \ddot \Psi(u)+\Omega^2 \Psi(u)=0
\end{align}
with the boundary conditions $\Psi(0)=0$ and $\dot \Psi(0)=1$. We finally obtain
\begin{align}
    G(t)=\sqrt{\frac{\alpha \Omega}{\sinh \Omega}}.
\end{align}


\begin{thebibliography}{10}{\color{red} }
\bibitem{caldeira1} A. O. Caldeira and A. J. Leggett, Ann. Phys. 149, 374 (1983).

\bibitem{caldeira2} A. O. Caldeira and A. J. Leggett, Physica A 121, 587 (1983).

\bibitem{caldeira3} A. O. Caldeira \textit{An Introduction to Macroscopic Quantum Phenomena and Quantum Dissipation} (Cambridge University Press, 2014).

\bibitem{Weiss} U. Weiss, \textit{Quantum Dissipative Systems} 3rd ed. (World Scientific, 2008).

\bibitem{Breuer} H.-P. Breuer and F. Petruccione, \textit{The Theory of Open Quantum System} (Oxford University Press, 2002).

\bibitem{Durao} L. M. M. Dur\~ao and A. O. Caldeira, Phys. Rev. E 94, 062147 (2016).

\bibitem{Oliveira} M. J. de Oliveira Phys. Rev. E 94, 012128 (2016).

\bibitem{Tsallis1} C. Tsallis, J. of Stat. Phys. 52 (1-2), 479-487.

\bibitem{Renyi} A. R\' {e}nyi, On Measures of Entropy and Information, Proceedings of the Fourth Berkeley Symposium on Mathematical Statistics and Probability, Volume 1: Contributions to the Theory of Statistics.

\bibitem{Tirnakli} U. Tirnakli and E. P. Borges, The standard map: From Boltzmann-Gibbs statistics to Tsallis statistics, Sci. Rep. 6, 23644 (2016).

\bibitem{Moyano} L. G. Moyano, C. Tsallis, and M. Gell-Mann, Numerical indications of a q-generalised central limit theorem, Europhys. Lett. 73, 813 (2006).

\bibitem{Tsallis2} C. Tsallis, Nonextensive statistics: Theoretical, experimental and computational evidences and connections, Braz. J. Phys. 29, 1 (1999).

\bibitem{Foster}  D. Forster, \textit{Hydrodynamics Fluctuations, Broken Symmetries and Correlation functions} (Benjamin, New York, 1975).

\bibitem{Zwanzig} R. Zwanzig, \textit{Nonequilibrium Statistical Mechanics} (Oxford University Press, 2001).

\bibitem{LeBellac} M. Le Bellac, F. Mortessagne and G. G. Batrouni, \textit{Equilibrium and Non-equilibrium Statistical Thermodynamics} (Cambridge University Press 2004).

\bibitem{Mukamel} S. Mukamel, \textit{Principles of Nonlinear Optical Spectroscopy} (Oxford university Press 1995)

\bibitem{Fox} M. Fox, \textit{Quantum Optics, An Introduction} (Oxford University Press 2006)

\bibitem{HMF1} P. Talkner and P. H\"anggi arXiv:1911.11660 [cond-mat.stat-mech] 27 Nov 2019.

\bibitem{HMF2} S. Hilt, B. Thomas, and E. Lutz Phys. Rev. E 84, 031110, Published 7 September 2011.

\bibitem{HMF3} M. Campisi, P. Talkner, and P. H\"anggi Phys. Rev. Lett. 102, 210401 (2009).

\bibitem{Landau} L. D. Landau and E. M. Lifshitz, \textit{Statistical Physics} Vol. 5 of the \textit{The Course of Theoretical Physics}.

\bibitem{Salinas} S. Salinas, \textit{Introduction to Statistical Physics} (Springer 2001).

\bibitem{Spohn} H. Spohn, J. of Math. Phys. 19, 1227 (1978); doi: 10.1063/1.523789.

\bibitem{RevTerm} R. Kosloff, Entropy 2013, 15, 2100-2128; doi:10.3390/e15062100.

\bibitem{Deffner} S. Deffner and E. Lutz, Phys. Rev. Lett 107, 140404 (2011).

\bibitem{Deffner2} S. Deffner, Euro. Phys. Lett., 103 (2013) 30001 www.epljournal.org doi: 10.1209/0295-5075/103/30001.

\bibitem{Aurell} E. Aurell and R. Eichhorn, New J. Phys. 17 (2015). 065007.

\bibitem{Theo} Th. M. Nieuwenhuizen and A. E. Allahverdyan, Phys. Rev. E 66, 036102 (2002). doi: 10.1103/PhysRevE.66.036102

\end{thebibliography}

\begin{thebibliography}{10}

\bibitem{caldeira3-S} A. O. Caldeira \textit{An Introduction to Macroscopic Quantum Phenomena and Quantum Dissipation} (Cambridge University Press, 2014).

\bibitem{Weiss-S} U. Weiss, \textit{Quantum Dissipative Systems} 3rd ed. (World Scientific, 2008).

\bibitem{Breuer-S} H.-P. Breuer and F. Petruccione, \textit{The Theory of Open Quantum System} (Oxford University Press, 2002).

\end{thebibliography}
\end{document}